
\font\btenbf=cmbx10 scaled\magstep1

\def \secth#1 {\goodbreak\noindent #1 \vskip 0.1truein}
\def \sectref {\goodbreak\noindent \vskip 0.1truein}

\def \ha {{\textstyle {1\over 2}}}

\newcount\sect
\newcount\no

\def \sectincr {\global \advance \sect by1}
\def \numb {\global \advance \no by1 \eqno(\number \number \no)}
\def \numba {\global \advance \no by1 {(\number \number \no)}}

\def \sectA {\global \let\sect=A}
\def \sectB {\global \let\sect=B}

\def \numbapp {\global \advance \no by1 \eqno(\sect \number \no)}

\def\pmb#1{\setbox0=\hbox{#1}%
\kern-.025em\copy0\kern-\wd0
\kern.05em\copy0\kern-\wd0
\kern-.025em\raise.0433em\box0}

\def \lg {Landau-Ginzburg\ }
\def \temp {temperature\ }
\def \temps {temperatures\ }
\def \at {\alpha_T}

\def\doublespace{\baselineskip=\normalbaselineskip
\multiply\baselineskip by 2 \divide\baselineskip by 1}

\magnification=1200
\doublespace

In the mean field description of
superconductors in a magnetic field
due to Abrikosov [1], the zeros of the
superconducting order parameter (vortices) form a triangular lattice
in the low temperature ordered phase.
This phase has a broken translational symmetry
which reflects long range positional order, and
phase coherence or off-diagonal long range order (ODLRO).
The conventional theoretical picture [2] supposes that the two-dimensional
Abrikosov vortex lattice
will undergo a continuous phase transition at a melting \temp $T_M$,
due to Kosterlitz-Thouless (KT)
unbinding of paired dislocations in the 2D crystal [3].
However, it has been argued [4] that ODLRO in the Abrikosov state is destroyed
by phase fluctuations due to thermal excitations of the lattice.
Assuming that the loss of ODLRO is accompanied by the loss of positional order
over the same correlation length scale $\xi$,
the implication then is that only
the disordered, vortex liquid state, will exist above zero \temp [5].

There have been various experimental papers
which claim to support the KT melting scenario for thin films [6-8].
These all used different
techniques, with changes in a measured quantity, such as resistivity or
ac penetration depth, near the expected phase boundary given as
evidence for the melting transition.
The resistivity which falls rapidly towards zero as the
temperature is lowered below a certain point
is usually interpreted as being due to the formation
of a flux lattice.
However, this rapid change can also be explained
as a consequence of
increasingly long
relaxation times which follow an Arrhenius law as the temperature is
lowered [7]. Furthermore, in  certain
regions of the H-T phase diagram for BSCCO crystals, the high
anisotropy make the separate layers behave as decoupled 2D systems.
Here the
resistivity drop can also be explained by an activated process [9] without
any need to invoke a phase transition.

Previous Monte Carlo (MC) studies of the 2D vortex system [10-13] have produced
conflicting results. Te\^sanovi\'c and Xing [10] find evidence for a
transition,
Kato and Nagaosa [11] and Hu and MacDonald [12]
conclude that
the vortex lattice melts through a {\it first order} phase transition
(for which there is no experimental support whatsoever),
while
O'Neill and Moore [13] find no evidence of
a vortex lattice at nonzero \temps.

This study extends the results of [13] by encompassing much larger systems and
longer MC sampling times (made possible by a much improved algorithm), but
nevertheless comes to the same conclusions, i.e. the {\it absence} of
any phase transition at finite temperatures.
In addition we can relate the relaxation times which we find in the system at
low temperatures to the behavior of the resistivity as seen in Ref. [9].

Our starting point is the \lg free energy functional of the complex
order parameter $\psi({\bf r})$,
which can be written as [13]
$$F[\psi({\bf r})]/k_BT_c
=\int d^3{\bf r}\,\{\alpha(T)|\psi|^2+\beta(T)|\psi|^4/2+|{\bf D}\psi|^2/2\mu\}
,\numb$$
where ${\bf D}=-i\hbar\nabla-2e{\bf A}$, the mean field zero-field
transition temperature is $T_c$, and $\alpha(T)$, $\beta$, $\mu$,
are the usual phenomenological parameters.

The functional integral for the partition function can be approximated
by expanding the order parameter in a basis set
consisting of only the lowest Landau level (LLL) eigenstates $\varphi_p$,
of the
gauge invariant operator ${\bf D}^2/2\mu$, i.e.
$$\psi=Q\sum_{p=0}^{N}v_{p}\varphi_p,\numb$$
where $Q=(\Phi_0/\beta Bd)^{1/4}$, (with $d$ the film thickness,
$B$ the external magnetic field), and $N$ is the number of vortices [13].
The vortices are constrained to move on the surface of a sphere
of radius $R$ which is
imagined to contain a monopole of fixed strength at the center,
such that $4\pi R^2B=N\Phi_0$.
The flux quantum $\Phi_0$ is related to the magnetic length $l_m$
via $\Phi_0=2\pi l_m^2B$.
Henceforth, $l_m$, which is the characteristic distance between vortices,
is used as the measure of
length; for example in these units, $R=(N/2)^{1/2}$.
An advantage of using this spherical geometry is that the vortex system has
full rotational invariance - a feature absent in
the geometry used in Refs. [11,12].

We neglect fluctuations in the ${\bf A}$ field and choose the gauge and
eigenfunctions $\varphi_p$ to be the same as those in Ref. [13].
Within this LLL approximation, the free energy in Eq. (1), becomes
$$F[\lbrace v_p\rbrace]/k_BT_c=\at\sum_{p=0}^{N}|v_p|^2+
\ha N^{-1}\sum_{p=0}^{2N}\left|U_p\right|^2,\numb$$
where
$$U_p=\sum_{q=0}^{N}h_{p-q}h_qB^{1/2}(N-q+1,q+1)\Theta(p-q)\,v_{p-q}v_q\numb$$
with the effective \temp variable $\at=d\,Q^2(\alpha+eB\hbar/\mu)$,
$\Theta(p-q)$ is the
Heaviside step function,
$h_p=[(N+2)!p!/2\pi N(N-p)!]^{1/2}$,
and
$B(x,y)=(x-1)!(y-1)!/(x+y-1)!$ is the Beta function.
Large negative values of $\at$ correspond to low \temps, and mean field
behavior is recovered (at \lq zero temperature') in the limit
$\at\rightarrow -\infty$.
The LLL approximation is expected to be valid at large fields approaching
the upper critical field $H_{c2}$, which correspond to small values of
$\at$.

The partition function requires integration of
$\exp\lbrace -F[\lbrace v_p\rbrace]/k_BT_c\rbrace$ over
the set of complex coefficients $\{v_p\}$.
The form of Eqs.(3) and (4) permits an efficient MC
sampling procedure to be devised.

To examine vortex correlations, we evaluate a \lq structure factor'
defined on the sphere as
$$S^m_l={1\over N}\int d\Omega d\Omega'\,
\langle|\psi(\Omega)|^2|\psi(\Omega')|^2\rangle
Y^m_l(\Omega)Y^{m*}_l(\Omega'),\numb$$
where $\Omega$ is the solid angle, $Y^m_l$ are spherical harmonics,
and $<..>$ denotes thermal averaging.
A simpler expression can be calculated by assuming that in
thermal equilibrium, the structure factor is azimuthally invariant,
so that we may choose $S(l)\equiv S^0_l$, and this is given by
$$S(l)={1\over N}\langle\,|\sum_p |v_p|^2\, I_{pl}|^2\,\rangle,\numb$$
where
$I_{pl}=Nh_p^2\int d\Omega\,|\varphi_p(\Omega)|^2P_l(\cos\theta)\exp{(l^2/4)}$,
and $P_l$ is the Legendre polynomial.

The standard Metropolis algorithm has been used to sample the distribution
of the dynamical variables $\{v_p\}$, with their initial configuration
generated
randomly. Typically, the first ten percent of the total number of
Monte Carlo steps (MCS) used in each run
are discarded to
allow the system to equilibrate, and
similarly, measurements from which averages are produced,
are taken every ten MCS.
System sizes of $N=120,160,200$, and $400$ were studied.

Some physical quantities such as the entropy (proportional
to $N^{-1}\sum_p\langle|v_p|^2\rangle$)
reach apparent
equilibrium very rapidly (on the order of $5\times 10^3$ MCS), and differs only
by about one percent from the largest system ($N=400$) to the smallest
($N=120$). In comparison, much longer sampling times are needed to accurately
determine the structure factor $S(l)$ of Eq. (6), ranging from $8\times 10^5$
to $2\times 10^6$ MCS.
In particular, for low \temps ($\at<-6$), at least $10^6$ MCS
are required to obtain reasonably stable data.
We will return to a more detailed discussion of the relaxation time scales
later.

First, we discuss the possibility of a first order transition.
Numerical simulations in the case of 2D systems
have always produced conflicting
results when searching for KT crystal melting [3]. Conclusions are often
heavily dependent on factors such as (dislocation) core energies and are highly
sensitive to boundary conditions. It is always difficult to do
simulations for
sufficiently large systems to minimize boundary and finite size effects.

Following Refs. [11-12], we have searched for signs of hysteresis behavior
in the energy and entropy, but without success.
It is useful
to investigate the probability distribution of the energy P(E),
when attempting to
interpret the nature of a phase transition [15]. Fig.1a is a plot of
this distribution, which shows no sign of the \lq double peak' expected for
a first order transition at the transition temperature
reported in [11] and
[12], corresponding to $\at\simeq -10$, for up to $N=400$.
Only a single peak is observed (at all \temps surrounding $\at=-10$) which
narrows with system size at a fixed \temp. In the asymptotic limit, it is
expected that the width at half height should tend to zero like $N^{-1/2}$ as
$N\rightarrow\infty$, and this is seen in Fig.1b.
Both of the studies [11] and [12] employ
quasi-periodic boundary conditions on the plane,
such that the ground state is {\it constructed} to be a lattice.
A possible reason then, for the appearance of a double peak, is that
as the correlation length $\xi$ over which crystal order exists
becomes of the order of the system size L, as it will for sufficiently low
temperature,
their system may easily
fall into an apparent low energy state.  However, at fixed \temp, this
situation might change as $L$ tends to infinity, although the crossover might
only happen at very large system sizes.
Furthermore, there is no sign of a first order transition in any experimental
data that we are aware of.

The structure factor of Eq. (6) as the \temp is lowered acquires peaks
at values of $l$ corresponding to the reciprocal lattice vectors $G=|{\bf G}|$
of the triangular lattice.
For a triangular
lattice, the first peak is at $G\simeq 2.694$, in terms of the normalised
\lq momentum' $l/R$.
The scaling argument stemming from a zero temperature transition [5,13]
predicts that the peak heights in the structure factor of the disordered
or liquid phase, should scale as
$S(G)\sim\xi^4\sim|\at|^4$.
We have determined $S(G)$ by extrapolating between two measurements
of $S(l)$ taken at $l/R$ values on either side of $l=G$.
That $\xi$ varies as $|\at|$ as $\at\rightarrow -\infty$, was established
in Ref. [13], and this has  been reinforced by our own results [5].
In particular, the peak widths, which vary as $1/\xi$, decrease
as $|\at|^{-1}$ as $\at\rightarrow -\infty$.

Turning now to the dynamics,
the running
value of the structure factor $S(l)$ at the reciprocal lattice, $l=G$,
was investigated
systematically as a function of time, and we shall call this function $S(G,t)$.
This particular value of $l$ was investigated in most detail as for this
wavevector, the \lq critical slowing down' associated with the
zero \temp phase transition, is most pronounced and hence
its study determines the longest relaxation time for the system.
We find that sometimes,
there appears to be more than one characteristic time scale. This is manifest
by a local (in time) minimum in the value of $S(G,t)$, which is different
to the eventual true equilibrium value. However, the initially relaxed local
value of $S(G,t)$ only changes by a few percent
as it \lq creeps' to the next locally stable value, as seen in Fig.2a.
Any local minima clearly correspond
to metastable states of the vortex system,
the first of which might be reached after a relatively short time
($<5\times 10^5$ MCS). In cases where even the initial relaxation time is
long ($>5\times 10^5$ MCS), or when $S(G,t)$ slowly tends to another metastable
state, we believe that the underlying mechanism for the longer timescales
involves creating and moving a dislocation
through correlated crystalline regions of the vortex system.
The energy of such a dislocation [3] within a \lq lattice' of
size $\xi$ is given by
$E_{\rm dis}\sim c_{66}dl_m^2\log(\xi/l_m)$,
where $c_{66}$ is the shear modulus.
Hence the activated relaxation time $\tau\sim\exp(E_{\rm dis}/k_BT)$ can be
written as
$$\tau=\tau_0\,\exp\{A|\at |^2\log(C|\at|)\},\numb$$
where $\tau_0$, $A$, and $C$ are constants, and we have used the
relations $|\at|\sim\xi$,
$\at^2\sim c_{66}dl_m^2/k_BT_c$ [13].

The value of $\tau$ has been extracted from the MC data by fitting
to plots of $S(G,t)$ at times {\it after} any initial relaxation,
either growing $\sim (1-e^{-t/\tau '})$
or decaying $\sim e^{-t/\tau '}$ exponentials tending
to an estimated (possibly local),
equilibrium value for $S(G,t)$. This procedure is intended to
measure the long timescale
so that in cases where $S(G,t)$
drifts away from the initially relaxed state at say time $\tau_i$
to the next
local minimum over a fitted time of $\tau '$, we take the measured relaxation
time to be $\tau=\tau_i+\tau '$.
A more formal determination
of $\tau$ through a calculation of
the autocorrelation function, was not feasible given the
large relaxation times involved.

The plot in Fig.2b
shows that Eq. (7) fits the data for $\at <-4$. The points for $N=120$
begin to
bend below the straight line at very low temperatures, or large $|\at|$, and
this is probably due to finite size effects (as can be seen from the
trend for the larger systems), when the correlation length $\xi$
becomes comparable with the linear system size $R$.

Yazdani et al [8], argued that by examining ac resistivity at high
frequencies it is possible to probe superconducting films on length
scales which are not affected by disorder.
The samples used were MoGe films, and the large decades of
response frequency covered (over a small temperature range)
imply a spread of relaxation times much wider than that
obtained from our MC simulations.
This is because
disorder does dominate in the low frequency regime (corresponding to
length scales larger than the
Larkin-Ovchinikov length),
at least in these conventional superconducting films.
In the case of
BSCCO crystals, where disorder is less important,
high fields applied perpendicular to the Cu-O
layers make
the layers decouple and hence act as 2D systems over a portion of the
H-T phase diagram [16],
providing a more
relevant comparison with our results.
Resistivity measurements at various fields as a function of temperature
below $T_c$,
show a gradual drop, typically exhibiting a \lq knee',  as the
temperature is lowered.
Below this \lq knee', the resistivity can be explained as a field dependent
activation process [9]. The relaxation process of Eq. (7) which is
the so called \lq plastic' timescale [16], is of this same Arrhenius
form.
Taking the resistivity to be inversely proportional to the relaxation time
of Eq. (7),
i.e. $\rho\sim 1/\tau$, it can be shown that our results match qualitatively
the data of ${\rm Brice\tilde no}$ et al [9] over a temperature range from
$\at\simeq -5$ to $-8$, corresponding to the data points just below
the knee.
At lower temperatures, disorder is
expected to be important again, leading to processes like TAFF which have
timescales that dominate
over the \lq plastic' ones [16].
However, we have not been able to obtain a quantitatively accurate fit as
it appears that the resistivity
data in ${\rm Brice\tilde no}$ et al does not belong
to the 2D scaling regime; in other words the measurements at different
fields do not collapse onto a single curve as a function of $\at$, implying
that the data is outside the LLL description.
(This is not surprising, given for instance, the small region in the H-T
diagram
which fit the 2D scaling regime in magnetisation data for BSCCO [17]).
It would be useful therefore, if a systematic investigation of the
resistivity was made in the regime where the layers are decoupled and
where the LLL approximation is valid. This would provide a direct
comparison with the simulation results reported here.

We would like to thank Y. Kato, N. Nagaosa, Jun Hu, A. Yazdani,
A.D. Rutenberg and M.A. Pettifer for helpful discussions.
The numerical simulations were performed with DEC Alpha 3000/400 workstations
and on a Cray EL98.
This work was supported through a SERC grant.

\vskip 1in
{\btenbf Figure captions}

FIG. 1. (a) The energy distribution curve, $P(E)$ plotted against
$E/E_{MF}$ where $E_{MF}=-N|\at|^2/2\beta_A$ is the mean field energy
(with $\beta_A\simeq 1.159$ the Abrikosov factor), at $\at=-10$ for $N=400$,
showing only a single peak.
The lack of a double peak suggests the absence of a first order phase
transition.
(b) The peak width at half height W, as a
function of $N^{-1/2}$. The straight line extrapolates to zero, indicating
the asymptotic regime has been reached.

FIG. 2. (a) Running value of $S(G,t)$ as a function of MC run time, for
$N=200$ at $\at=-8$
showing an initial relaxation to a local metastable state after
$8\times 10^5$ MCS, and then the subsequent \lq creep' to another metastable
state after $1.3\times 10^6$ MCS in total.
(b) The logarithm of the relaxation time for various system sizes plotted
against $|\at|^2\log(|\at|)$. A linear fit is obtained for temperatures
lower than $\at\simeq -5$, verifying Eq. (7). The falloff at larger $|\at|$
are due to finite size effects.

\vskip 5in
\sectref{{\btenbf References}}

\vskip 0.1truein
[1] A.A. Abrikosov, Zh. Eskp. Teor. Fiz. {\bf 32}, 1442 (1957).

[2] S. Doniach and B. A. Hubermann, Phys. Rev. Lett. {\bf 42}, 1169 (1979);

\hskip 0.2in D. S. Fisher, Phys. Rev. B {\bf 22}, 1190 (1980).

[3] K.J. Strandburg, Rev. Mod. Phys. {\bf 60}, 161 (1988).

[4] M.A. Moore, Phys. Rev. B {\bf 45}, 7336 (1992).

[5] M.A. Moore and H.H. Lee (to be published)

[6] P.L. Gammel, A.F. Hebard, and D.J. Bishop, Phys. Rev. Lett. {\bf 60},
144 (1988).

[7] P. Berguis, A.L.F. van der Slot, and P.H. Kes, Phys. Rev. Lett. {\bf 65},
2583 (1990);

\hskip 0.2in P. Berghuis and P.H. Kes, Phys. Rev. B {\bf 47}, 262 (1993).

[8] A. Yazdani, W.R. White, M.R. Hahn, M.Gabay, M.R. Beasley, and A.Kapitulnik,

\hskip 0.2in Phys. Rev. Lett. {\bf 70}, 505 (1993).

[9] G. ${\rm Brice\tilde no}$, M.F. Crommie, and A.Zettl, Phys. Rev. Lett.
{\bf 66}, 2164 (1991);

\hskip 0.2in T.T.M. Palstra, B.Batlogg, L.F. Schneemeyer, and J.V. Waszczak,

\hskip 0.2in Phys. Rev. Lett. {\bf 61}, 1662 (1988).

[10] Z. Te\^sanovi\'c and L. Xing, Phys. Rev. Lett. {\bf 67}, 2729 (1991).

[11] Y. Kato and N. Nagaosa, Phys. Rev. B {\bf 47}, 2932 (1993);

\hskip 0.2in Y. Kato and N. Nagaosa, Phys. Rev. B {\bf 48}, 7383 (1993).

[12] J. Hu and A.H. MacDonald, Phys. Rev. Lett. {\bf 71}, 432 (1993).

[13] J.A. O'Neill and M.A. Moore, Phys. Rev. B {\bf 48}, 374 (1993);

\hskip 0.2in J.A. O'Neill and M.A. Moore,
Phys. Rev. Lett. {\bf 48}, 374 (1992).

[14] G.J. Ruggeri and D. J. Thouless, J. Phys. F {\bf 6}, 2063 (1976).

[15] J.Lee and J.M. Kosterlitz, Phys. Rev. Lett. {\bf 65}, 137 (1990).

[16] G. Blatter, M.V. Feigel'man, V.B. Geshkenbein, A.I. Larkin, and
V.M. Vinokur,

\hskip 0.2in preprint.

[17] N. Wilkin and M. A. Moore, Phys. Rev. B {\bf 48}, 3464 (1993).

\end